\documentclass[twocolumn,epjc3]{svjour3}

\usepackage{amsmath}
\usepackage{euscript,amssymb,epsfig}
\usepackage{graphicx}
\usepackage{amsfonts,latexsym}

\RequirePackage{graphicx}

\newcommand{\be}[1]{\begin{equation}\label{#1}}
\newcommand{\ee}{\end{equation}}
\newcommand{\ba}[1]{\begin{eqnarray}\label{#1}}
\newcommand{\ea}{\end{eqnarray}}
\newcommand{\rf}[1]{(\ref{#1})}
\newcommand{\nn}{\nonumber}

\journalname{Eur. Phys. J. C}

\begin{document}

\title{Many-body problem in Kaluza-Klein models with toroidal compactification}

\author{Alexey Chopovsky\thanksref{e1,addr1,addr2} \and Maxim Eingorn\thanksref{e2,addr2,addr3,addr4} \and Alexander Zhuk\thanksref{e3,addr2}}

\thankstext{e1}{e-mail: a.chopovsky@yandex.ru}

\thankstext{e2}{e-mail: maxim.eingorn@gmail.com}

\thankstext{e3}{e-mail: ai.zhuk2@gmail.com}

\institute{Department of Theoretical Physics, Odessa National University, Dvoryanskaya st. 2, Odessa 65082, Ukraine\label{addr1}
           \and
           Astronomical Observatory, Odessa National University, Dvoryanskaya st. 2, Odessa 65082, Ukraine \label{addr2}
           \and
Department of Theoretical and Experimental Nuclear Physics, Odessa National Polytechnic University, Shevchenko av. 1, Odessa 65044, Ukraine \label{addr3} \and
North Carolina Central University, CREST and NASA Research Centers, Fayetteville st. 1801, Durham, North Carolina 27707, U.S.A. \label{addr4}}

\date{Received: date / Accepted: date}

\maketitle

\begin{abstract}
In this paper, we consider a system of gravitating bodies in Kaluza-Klein models with toroidal compactification of extra dimensions. To simulate the
astrophysical objects (e.g., our Sun and pulsars) with energy density much greater than pressure, we suppose that these bodies are pressureless in the
external/our space. At the same time, they may have nonzero parameters $\omega_{(\bar\alpha -3)} \, (\bar\alpha =4,\ldots ,D) $ of the equations of state  in
the extra dimensions. We construct the Lagrange function of this many-body system for any value of $\Sigma =\sum_{\bar\alpha} \omega_{(\bar\alpha -3)}$.
Moreover, the gravitational tests (PPN parameters, perihelion and periastron advances) require negligible deviation from the latent soliton value $\Sigma
=-(D-3)/2$. However, the presence of pressure/tension in the internal space results necessarily in the smearing of the gravitating masses over the internal
space and in the absence of the KK modes. This looks very unnatural from the point of quantum physics.
\end{abstract}



\section{\label{sec:1}Introduction}

\setcounter{equation}{0}

The idea of multidimensionality of our Universe demanded by the theories of unification of the fundamental interactions is one of the most breathtaking ideas
of theoretical physics. It takes its origin from the pioneering papers by Th. Kaluza and O. Klein \cite{KK}, and now the most self-consistent modern theories
of unification such as superstrings, supergravity and M-theory are constructed in spacetimes with extra dimensions (see, e.g., \cite{Polchinski}). Different
aspects of the idea of multidimensionality are intensively used in numerous modern articles.

Therefore, it is important to find experimental evidence for the existence of the extra dimensions. For example, one of the aims of Large Hadronic Collider
consists in detecting of Kaluza-Klein (KK) particles which correspond to excitations of the internal spaces (see, e.g., \cite{KKparticles}). Such excitations
were investigated in a lot of articles (see, e.g., the classical papers \cite{SS,Nieu,Kim}). Quite recently, KK particles were considered, e.g., in the papers
\cite{KLZ,BD}.

On the other hand, if we can show that the existence of the extra dimensions is contrary to observations, then these theories are
prohibited.

Much work was done in this direction including the models with toroidal compactification. Obviously, any gravitational theory modified with respect to the
General Relativity (GR) can result in some observable deviations form GR. A number of papers were devoted to the search of such deviations. For example, the
nonrelativistic gravitational potentials in these theories can be different from the Newtonian potentials \cite{ADD1,ADD2,KeSf,BarvSol,CallBurg,EZ1,EZ2}. In
principle, this difference can be experimentally observed \cite{Hoyle}. Parameterized Post-Newtonian formalism is a powerful tool for the determination of
gravitational theories consistent with experiments \cite{XuMa,EZ3}.

The relation with particle physics is another important point of KK models. It was shown that multidimensional models can give a reasonable explanation of the
hierarchy problem \cite{ADD1,ADD2}. Then, it was indicated that such framework can be embedded in the string theory \cite{AntA-HDD}. On the other hand, the
interaction between KK states and ordinary matter can result in new observable channels of reactions \cite{ADD1,ADD2,AntA-HDD,Giu,Lykk1,Lykk2,Hann}.

In our previous papers \cite{EZ3,EZ4,EZ5} devoted to KK models with toroidal compactification of the extra dimensions, we have shown that gravitating masses
should have tension in the internal space to be in agreement with gravitational experiments in the Solar system. For example, black strings/branes with the
parameter $\omega =-1/2$ of the equation of state in the internal space satisfy this condition. For this value of $\omega$, the variations of the internal
space volume are absent \cite{EZ6}. In the dust-like case with $\omega =0$, such variations generate the fifth force, that leads to contradictions with the
experimental data.
%in the case of model with the pressureless equations of state in the internal space.
%Hence, for individual gravitating bodies, we have shown in which cases and why the KK models with toroidal compactification are in agreement %with the
%observations. Is it sufficient to claim that these models are viable?

It is worth noting that black strings/branes generalize the known Schwarzschild solution to the multidimensional case (see, e.g., \cite{TF,TZ,HO,TK} and the
corresponding literature therein). Obviously, any multidimensional theory should have such solutions, as they must correspond to the observed astrophysical
objects. Black strings/branes have toroidal compactification of the internal spaces. This compactification type is the simplest among the possible ones.
However, it makes sense to investigate such models because they may help to reveal new important properties for more physically reliable multidimensional
models. The ADD model \cite{ADD1} presents a good example of it. Even if the authors use the localization of the Standard model fields on a brane, they explore
the toroidal compactification of the internal space to get the relation between the multidimensional and four-dimensional gravitational constants \cite{ADD2}.
That gives a possibility to solve the hierarchy problem and to introduce the notion of large extra dimensions. We will not use the brane approach for our model
remaining within the standard Kaluza-Klein theory. However, even in this case the large extra dimensions can be achieved for KK models with toroidal
compactification \cite{EZ2}.

The main purpose of this paper is to construct the Lagrange function for a many-body system in the case of models with toroidal compactification. We need such
theory e.g. to calculate the formula for advance of periastron in the case of a binary system. The measurement of this advance for the pulsar PSR B1913+16 was
performed with very high accuracy. Therefore, such measurements can be a very good test for gravitational theories. From our previous papers \cite{EZ3,EZ4,EZ5}
we know that gravitating bodies should have pressure/tension in the extra dimensions to satisfy the observable data for the deflection of light and the
experimental restrictions for the parameterized post-Newtonian parameter (PPN) $\gamma$. In this regard, the question arises about the possibility of building
a many-body Lagrange function in the presence of pressure/tension in the extra dimensions. To answer this question, we need the  metrics components $g_{00}$ up
to $O(1/c^4)$, $g_{0\alpha}$ up to $O(1/c^3)$ and $g_{\alpha\beta}$ up to $O(1/c^2)$. It is worth noting that for the expressions of the deflection of light
and PPN parameter $\gamma$, it is sufficient to calculate the metrics coefficients up to $O(1/c^2)$. Obviously, the agreement with observations up to
$O(1/c^2)$ does not guarantee the agreement up to $O(1/c^4)$. Hence, we calculate the metrics coefficients in the required orders $1/c$. We demonstrate that
the many-body Lagrange function can be constructed for any value of $\Sigma$ where $\Sigma$ is a sum of the parameters of the equations of state in the extra
dimensions. We demonstrate that the gravitational tests (PPN parameter $\gamma$, and perihelion/periastron advance) allow very small deviation from the latent
soliton value $\Sigma =-(D-3)/2 \neq 0$. We prove that nonzero $\Sigma$ leads necessarily to the uniform smearing of the gravitating masses over the internal
space. However, uniformly smeared gravitating bodies cannot have  excited KK states (KK particles). As we mentioned above, KK particles were recently
considered in the papers \cite{KLZ,BD}. Here, the metric and form-field perturbations are studied without taking into account the reason of such fluctuations.
Our present analysis clearly shows that the inclusion of the matter sources, being responsible for the perturbations, imposes strong restrictions on the model,
e.g., leading to the absence of KK particles. Until now, KK particles were not detected in experiments at LHC. So, it looks tempting to interpret their absence
in the light of our paper (i.e. due to the smearing of the gravitating particles over the internal space). However, the absence of KK particles looks rather
unnatural from the point of quantum mechanics and statistical physics (see below). Therefore, in our opinion, this is a big disadvantage of the Kaluza-Klein
models with the toroidal compactification.

The paper is structured as follows. In Sec. 2, we obtain the $1/c^2, 1/c^3$ and $1/c^4$ correction terms to the metric coefficients for the considered
many-body system. In Sec. 3, we demonstrate that gauge conditions lead to the uniform smearing of the gravitating bodies over the extra dimensions. The
Lagrange function for the many-body system is constructed in the Sec. 4. The formulas for PPN parameters $\beta, \gamma$ and perihelion and periastron advances
are calculated in Section 5. These formulas allow us to obtain  experimental constraints on the parameters of the model. The main results are summarized in
concluding Sec. 6.

%%%%%%%%%%%%%%%%%%%%%%%%%%%%%%%%%%%%%%%%%%%%%%%%%%%%%%%%%%%%%%%%%%%%%%%%%%%%%%%%%%%
\section{\label{sec:2}Metric coefficients in the weak field approximation}

To construct the Lagrange function of a system of $N$ massive bodies in $(D+1)$-dimensional spacetime, we define first the nonrelativistic gravitational field
created by this system. To do it, we need to get the metric coefficients in the weak field limit. The general form of the multidimensional metrics is
%%%%%
\be{2.1}
ds^2=g_{ik}dx^idx^k=g_{00}\left(dx^0\right)^2+2g_{0\mu}dx^0dx^{\mu}+g_{\mu\nu}dx^{\mu}dx^{\nu}\, ,
\ee
%%%%%
where the Latin indices $i,k = 0,1,\ldots ,D$ and the Greek indices  $\mu ,\nu = 1,\ldots ,D$.
%$D$ is the total number of
%spatial dimensions.
%and $d$ is the number of extra spatial dimensions.
We make the natural assumption that in the case of the absence of matter sources the spacetime is Minkowski spacetime: $g_{00}=\eta_{00}=1$,
$g_{0\mu}=\eta_{0\mu}=0$, $g_{\mu\nu}=\eta_{\mu\nu}=-\delta_{\mu\nu}$. In our paper, we consider in detail the case where the extra dimensions
%may
have the topology of tori. In the presence of matter, the metrics is not the Minkowskian one, and we investigate it in the weak field limit. It means that the
gravitational field is weak and velocities of test bodies are small compared with the speed of light $c$. In the weak field limit the metrics is only slightly
perturbed from its flat spacetime value. We will define the metrics \rf{2.1} up to $1/c^2$ correction terms. Because the coordinate $x^0=c t$, the metric
coefficients can be expressed as follows:
%%%%%%
\ba{2.2} &{}&g_{00}\approx1+h_{00}+f_{00}\, ,\; \; g_{0\mu}\approx h_{0\mu}+f_{0\mu}\, ,\; \; \nn\\
&{}&g_{\mu\nu}\approx-\delta_{\mu\nu}+h_{\mu\nu}\, , \ea
%%%%%%
where $h_{ik}\sim O(1/c^2), f_{00}\sim O(1/c^4)$  and $f_{0\mu}\sim O(1/c^3)$. In particular, $h_{00} \equiv 2\varphi /c^2$ where
$\varphi $ is the nonrelativistic gravitational potential.
To get these correction
terms, we should solve (in the corresponding orders of $1/c$) the multidimensional Einstein equation
%%%%
\be{2.3}
R_{ik}=\frac{2S_D\tilde G_{\mathcal{D}}}{c^4}\left(T_{ik}-\frac{1}{D-1}g_{ik}T\right)\, ,
\ee
%%%%
where $S_D=2\pi^{D/2}/\Gamma (D/2)$ is the total solid angle (the surface area of the $(D-1)$-dimensional sphere of the unit radius), $\tilde G_{\mathcal{D}}$
is the gravitational constant in the $(\mathcal{D}=D+1)$-dimensional spacetime. We consider a system of $N$ discrete massive (with rest masses $m_p,\,
p=1,\ldots ,N$) bodies. We suppose that the pressure of these bodies in the external three-dimensional space is much less than their energy density. This is a
natural approximation for ordinary astrophysical objects such as our Sun. For example, in general relativity, this approach works well for calculating the
gravitational experiments in the Solar system \cite{Landau}. In the case of pulsars, pressure is not small but still much less than the energy density, and the
pressureless approach was used in General Relativity to get the formula of the periastron advance \cite{Will2}. Therefore, the gravitating bodies are
pressureless in the external/our space. On the other hand, we suppose that they may have pressure in the extra dimensions. Therefore, nonzero components of the
energy-momentum tensor of the system can be written in the following form:
%%%%%%
\ba{2.4}
&{}&T^{ik}=\tilde\rho c^2u^iu^k, \quad i,k=0,\ldots ,3\, ,\\
&{}&\label{2.5} T^{i\bar\alpha}=\tilde\rho c^2 u^iu^{\bar\alpha}\, ,\quad
i=0,\dots ,3;\; \bar{\alpha}=4,\ldots ,D\, ,\\
&{}&\label{2.6} T^{\bar\alpha\bar\beta}=-p_{(\bar\alpha-3)}g^{\bar\alpha\bar\beta}+\tilde\rho c^2 u^{\bar\alpha}
u^{\bar\beta},\quad\bar{\alpha},\bar{\beta}=4,\ldots ,D\, ,\ea
%%%%%%
where the $(D+1)$-velocity $u^i=dx^i/ds$ and
%%%%%%
\be{2.7} \tilde\rho \equiv \sum_{p=1}^N\left[(-1)^Dg\right]^{-1/2} m_p\sqrt{g_{lm}\cfrac{dx^l}{dx^0}\frac{dx^m}{dx^0}}\delta({\bf x}-{\bf x}_p)\, , \ee
%%%%%%%
where ${\bf x}_p$ is a $D$-dimensional radius-vector of the $p$-th particle.
%%%%%
In what follows, the Greek indices $\alpha,\beta = 1,2,3$; $\bar\alpha,\bar\beta = 4,\ldots ,D$ and $\mu,\nu$ still run from 1 to $D$.
In the extra dimensions we suppose the equations of state:
%%%%%%
\be{2.8}
p_{(\bar\alpha-3)}=\omega_{(\bar\alpha-3)}\tilde\rho c^2\, .
\ee
%%%%%%
If all parameters $\omega_{(\bar\alpha-3)}=0$, then we come back to the model considered in our paper \cite{EZ3}. Here, massive bodies have dust-like equations
of state in all spatial dimensions. If all $\omega_{(\bar\alpha-3)}=-1/2$ (tension in the extra dimensions), then these equations of state correspond to black
strings (in the case of one extra dimension, i.e. $D=4$) and black branes (for $D>4$) \cite{TF,TZ,HO,TK}. If parameters satisfy the condition
$\sum\limits_{\bar\alpha}\omega_{(\bar\alpha-3)}\equiv\Sigma = - (D-3)/2$, then this case corresponds to latent solitons \cite{EZ5}. Obviously, black
strings/branes satisfy this condition.

Now, we will solve the Einstein equation \rf{2.3} in the corresponding orders of $1/c$. Obviously, for $\omega_{(\bar\alpha-3)}=0\, ,\bar\alpha =4,\ldots ,D$,
we should reproduce the results of the paper \cite{EZ3}. Because our calculations generalize the ones in \cite{EZ3}, we skip some evident details.

First, to get the metric correction terms of the order $O(1/c^2)$, the energy-momentum tensor components \rf{2.4}-\rf{2.6} are approximated as
%%%%%
\ba{3.1}
T^{00}&\approx& T_{00}\approx \rho c^2, \quad
T^{\bar\alpha\bar\beta}\approx T_{\bar\alpha\bar\beta}\approx \omega_{(\bar\alpha -3)}\, \rho c^2\, \delta_{\bar\alpha\bar\beta},\nn \\
T^{0\mu}&\approx& -T_{0\mu}\approx 0, \quad T^{\alpha\beta}\approx T_{\alpha\beta}\approx 0, \nn \\
T&=&T^{ik}g_{ik}\approx \rho c^2 (1-\Sigma)\, ,
\ea
%%%%%
where
%%%%%
\be{3.2}
\Sigma \equiv \sum\limits_{\bar\alpha=4}^D\omega_{(\bar\alpha-3)}
\ee
%%%%%
and we introduced the rest-mass density
%%%%%%
\be{3.3}
\rho ({\bf x}) = \sum_{p=1}^Nm_p\delta{({\bf x}-{\bf x}_p)}\, .
\ee
%%%%%%
Then,
from the Einstein equation we get
%%%%%%
\ba{3.6}
&{}&h_{00} =\frac{2\varphi({\bf x})}{c^2},\quad h_{0\mu}=0,\quad\\
&{}&\label{3.7} h_{\alpha\beta}=\frac{1-\Sigma}{D-2+\Sigma}\,\frac{2\varphi({\bf x})}{c^2}\delta_{\alpha\beta},\\
&{}&\label{3.8} h_{\bar\alpha\bar\beta}= \frac{ \omega_{(\bar\alpha-3)}(D-1) +1-\Sigma}{D-2+\Sigma}\, \frac{2\varphi({\bf
x})}{c^2}\delta_{\bar\alpha\bar\beta}\, ,\ea
%%%%%%
where the function $\varphi({\bf x})$ satisfies the $D$-dimensional Poisson equation
%%%%%
\be{3.9}
\triangle_D\varphi({\bf x})=2S_{D}\tilde G_{\mathcal{D}}\cfrac{D-2+\Sigma}{D-1}\,\rho({\bf x})\, .
\ee
%%%%%
We would remind that ${\bf x}$ is a $D$-dimensional radius-vector. It is worth noting that if $\omega_{(\bar\alpha-3)}=0,\; \forall \; \bar\alpha\;
\Rightarrow\; \Sigma =0$, then we reproduce the results of the paper \cite{EZ3}. On the other hand, if all $\omega_{(\bar\alpha-3)}=-1/2$, then
$h_{\bar\alpha\bar\beta}=0$ that should take place for black strings/branes \cite{EZ5}.

Next, we should obtain the $O(1/c^4)$ and $O(1/c^3)$ metric correction terms $f_{00}$ and $f_{0\mu}$, respectively. In this case, the energy-momentum
components read
%%%%%%
\ba{4.2}
&{}&T_{00}\approx\rho c^2 \left[1+\cfrac{\varphi}{c^2}\cfrac{3D-4+\Sigma}{D-2+\Sigma} +\cfrac{v^2}{2c^2}\right]\, ,\\
&{}&T_{0\mu}\approx -\rho c v^{\mu}\, ,\label{4.3}\\
&{}&T_{\alpha\beta}\approx \rho v^{\alpha}v^{\beta}\, ,\quad T_{\alpha\bar\beta}\approx \rho v^{\alpha}v^{\bar\beta}\, ,\label{4.4}\\
\label{4.5} &{}&T_{\bar\alpha\bar\beta}\approx \rho c^2\left\{\omega_{(\bar\alpha-3)}\delta_{\bar\alpha\bar\beta} \left[1 +\cfrac{\varphi}{c^2}
\right.\right.\nn\\
&{}& \times\left.\left.\cfrac{D-\Sigma-2\left[\omega_{(\bar\alpha-3)}(D-1)+1-\Sigma\right]}{D-2+\Sigma}\right.\right.\nn\\
&{}& -\left.\left.\cfrac{v^2}{2c^2}\right]+ \cfrac{v^{\bar\alpha}v^{\bar\beta}}{c^2}\right\}\, ,\ea
%%%%%%
and the trace
%%%%%%
\be{4.6}
T\approx
\rho c^2(1-\Sigma)+\rho\varphi\cfrac{(D-\Sigma)(1-\Sigma)}{D-2+\Sigma}+\rho(\Sigma-1)\frac{v^2}{2}\, .
\ee
%%%%%%%
Then, from the Einstein equation we get
%%%%%%
\ba{4.12} &{}&f_{00}({\bf x})=\cfrac{2}{c^4}\,\varphi^2({\bf x})
+\cfrac{2}{c^4}\sum_p\varphi_p({\bf x}-{\bf x}_p)\varphi'({\bf
x}_p)\nn \\
&+& \cfrac{1}{c^4}\cfrac{D-\Sigma}{D-2+\Sigma}\sum_p\varphi_p({\bf x}-{\bf x}_p)v_p^2\, .
\ea
%%%%%%%
and
%%%%%%
\be{4.14}
f_{0\mu}({\bf x})=-\cfrac{2}{c^3}\cfrac{D-1}{D-2+\Sigma}\sum_p\varphi_p({\bf x}-{\bf x}_p)v^\mu_p-\cfrac{1}{c^3}\cfrac{\partial^2f}{\partial t\partial x^\mu}\, ,
\ee
%%%%%%
where the function $f$ satisfies the following equation:
%%%%%%
\be{4.15}
\triangle_D f=\varphi({\bf x})\ .
\ee
%%%%%%

%%%%%%%%%%%%%%%%%%%%%%%%%%%%%%%%%%%%%%%%%%%%%%%%%%%%%%%%%%%%%%%%%%%%%%%%%%%%%%%%%%%
\section{\label{sec:5}Gauge conditions and smearing}

It should be noted that to calculate the Ricci tensor components in the corresponding orders of $1/c$, we used the standard (see, e.g., Eq. (105.10) in
\cite{Landau}) gauge condition
%%%%%
\be{5.1}
\partial_k\left(h^k_i-\cfrac{1}{2}\,h_l^l\delta_i^k\right)=0\, , \quad i,k =0,1,\ldots , D\, ,
\ee
%%%%%
where $h^k_i \equiv \eta^{km}h_{mi}$. Hence,
%%%%%
\be{5.2}
h^0_0=\eta^{00}h_{00}=h_{00}, \quad h^{\mu}_{\nu}=\eta^{\mu\kappa}h_{\nu\kappa}=-h_{\mu\nu}\, .
\ee
%%%%%
Therefore,
%%%%%%%
\ba{5.3}
h_0^0&=&\cfrac{2\varphi({\bf x})}{c^2}, \quad h_{\alpha}^{\beta}=-\cfrac{1-\Sigma}{D-2+\Sigma}\; \cfrac{2\varphi({\bf x})}{c^2}\delta_{\alpha}^{\beta},\\
h_{\bar\alpha}^{\bar\beta}&=&-\cfrac{\omega_{(\bar\alpha-3)}(D-1)+1-\Sigma}{D-2+\Sigma}\; \cfrac{2\varphi({\bf x})}{c^2}\delta_{\bar\alpha}^{\bar\beta}\, ,
\label{5.4} \\
h_l^l&=&\cfrac{2(\Sigma-1)}{D-2+\Sigma}\; \cfrac{2\varphi({\bf x})}{c^2}\, .\label{5.5}
\ea
%%%%%%%
Let us check that these solutions satisfy the condition \rf{5.1}. For $i=0$, we get immediately
%%%%%
\be{5.6}
\partial_k\left(h^k_0-\cfrac{1}{2}\,h_l^l\delta_0^k\right)=
\partial_0\left(h^0_0-\cfrac{1}{2}\,h_l^l\right)=0 +O\left(\frac{1}{c^3}\right)\, .
\ee
%%%%%
For $i=\beta$ we have
%%%%%%
\ba{5.7}
&{}&\partial_k\left(h^k_{\beta}-\cfrac{1}{2}\,h_l^l\delta_{\beta}^k\right)=
\partial_{\alpha}\left(h^{\alpha}_{\beta}-\cfrac{1}{2}\,h_l^l\delta_{\beta}^{\alpha}\right)\nn \\
&=&\left[-\cfrac{1-\Sigma}{D-2+\Sigma}+\cfrac{1-\Sigma}{D-2+\Sigma}\right]\cfrac{2}{c^2}\,\partial_{\beta}\varphi=0\, ,
\ea
%%%%%%
that is the condition is automatically satisfied. For $i=\bar\beta$ we obtain
%%%%%%
\ba{5.8} &{}&\partial_k\left(h^k_{\bar\beta}-\cfrac{1}{2}\,h_l^l\delta_{\bar\beta}^k\right)=
\partial_{\bar\alpha}\left(h^{\bar\alpha}_{\bar\beta}-\cfrac{1}{2}\,h_l^l\delta_{\bar\beta}^{\bar\alpha}\right)
\nn\\
&=&-\cfrac{\omega_{(\bar\beta-3)}(D-1)}{D-2+\Sigma} \cfrac{2}{c^2}\,\partial_{\bar\beta}\varphi=0\, . \ea
%%%%%%
In order to satisfy this condition, we should demand either $\omega_{(\bar\beta-3)}=0$ or $\partial_{\bar \beta} \varphi=0$. Because we consider the general
case $\omega_{(\bar\beta-3)}\neq 0$, we must choose the latter condition. Moreover, the gravitational tests require nonzero $\omega_{(\bar\beta-3)}$ (see Sec.
5). Therefore, the presence of nonzero pressure/tension in the extra dimensions results in the metric coefficients which do not depend on the coordinates of
the internal space, i.e. the gravitating masses should be uniformly smeared over the extra dimensions. In this case, the rest mass density \rf{3.3} should be
rewritten in the form: $\rho({\bf x})\rightarrow \rho({\bf r})=\sum_pm_p\delta({\bf r}-{\bf r}_p)/\prod_{\bar\alpha}a_{(\bar\alpha -3)}$, where ${\bf r}_p$ is
a three-dimensional radius vector of the $p$-th particle in the external space, $a_{(\bar\alpha -3)}$ are periods of the tori (i.e.
$\prod_{\bar\alpha}a_{(\bar\alpha -3)}$ is the volume of the internal space). Then, Eq. \rf{3.9} is reduced to the ordinary three-dimensional Poisson equation
%%%%%
\be{5.9}
\triangle_3\varphi({\bf r})=4\pi G_N\sum_{p}m_p\delta({\bf r}-{\bf r}_p)
\ee
%%%%%
with the solution
%%%%%%
\be{5.10}
\varphi({\bf r})=-\sum_{p}\frac{G_N m_p}{|{\bf r}-{\bf r}_p|}=\sum_p \varphi_p ({\bf r}-{\bf r}_p)\, ,
\ee
%%%%%%
where $G_N$ is the Newtonian gravitational constant:
%%%%%
\be{5.11}
4\pi G_N =\frac{2S_D(D-2+\Sigma)}{(D-1)\prod_{\bar\alpha}a_{(\bar\alpha -3)}} \tilde G_{\mathcal{D}}\, .
\ee
%%%%%
Hereafter, ${\bf r}, {\bf r}_p$ are radius vectors in three-dimensional external/our space.

In the case of the smearing, Eq. \rf{4.15} has the following solution
%%%%%%
\be{5.12}
f({\bf r})=-\cfrac{G_N}{2}\sum_pm_p|{\bf r}-{\bf r}_p|\, ,
\ee
%%%%%%
where, to get it, we used the well known equation $\triangle_3 r = 2/r$ in the three-dimensional flat space. Because
%%%%%%
\ba{5.13} &{}&\cfrac{\partial}{\partial t}\left(\cfrac{\partial|{\bf r}-{\bf r}_p|}{\partial x^{\alpha}}\right)=\cfrac{\partial}{\partial
t}\left(\cfrac{x^{\alpha}-x^{\alpha}_p}{|{\bf r}-{\bf r}_p|}\right)= \cfrac{1}{|{\bf r}-{\bf r}_p|^2}\nn\\
&\times&\left[-v^{\alpha}_p |{\bf r}-{\bf r}_p|-\cfrac{x^{\alpha}-x^{\alpha}_p}{|{\bf r}-{\bf r}_p|} \sum_{\beta}(x^{\beta}-x^{\beta}_p)(-v^{\beta}_p)\right]\,
, \ea
%%%%%%
we get for $f_{0\alpha}$:
%%%%%%
\be{5.14} f_{0\alpha} =\cfrac{G_N}{2c^3}\sum_p\cfrac{m_p}{|{\bf r}-{\bf r}_p|}\left(\frac{3D-2-\Sigma}{D-2+\Sigma}v^{\alpha}_p +n_p^{\alpha} ({\bf n}_p{\bf
v}_p)\right)\, , \ee
%%%%%%%
where we introduce the three-dimensional unit vector in the direction from the $p$-th particle to a point with the radius vector ${\bf r}$:
%%%%%
\be{5.15} n_p^{\alpha}=\cfrac{x^{\alpha}-x_p^{\alpha}}{|{\bf r}-{\bf r}_p|}\, , \ee
%%%%%
and $({\bf n}_p{\bf v}_p)=
\sum_{\beta}n_p^{\beta}v^{\beta}_p$.

It should be noted that, to get the formula \rf{4.14}, we used the following gauge condition:
%%%%%%
\be{5.16}
\cfrac{\partial f_0^\mu}{\partial x^\mu}-\cfrac{1}{2}\,\cfrac{\partial h_\mu^\mu}{\partial x^0} = 0\, ,
\ee
%%%%%%
where $f_0^{\mu}=\eta^{k\mu}f_{0k}=-f_{0\mu}$. In the case of smearing, this condition is reduced to
%%%%%
\be{5.17}
\cfrac{\partial f_0^\beta}{\partial x^\beta}-\cfrac{1}{2}\,\cfrac{\partial h_\mu^\mu}{\partial x^0} = 0\, ,
\ee
%%%%%
where we remind that $\alpha,\beta = 1,2,3$ and $\mu,\nu = 1,\ldots ,D$. Taking into account the following auxiliary equations:
%%%%%
\ba{5.18}
&{}&\cfrac{\partial}{\partial x^{\beta}}\cfrac{1}{|{\bf r}-{\bf r}_p|}=-\cfrac{n_p^{\beta}}{|{\bf r}-{\bf
r}_p|^2}\, ,\\
&{}&\sum_{\beta}\cfrac{\partial}{\partial x^{\beta}}\left(n_p^{\beta}\cfrac{({\bf n}_p{\bf v}_p)}{|{\bf r}-{\bf r}_p|}\right)=\cfrac{({\bf
n}_p{\bf v}_p)}{|{\bf r}-{\bf r}_p|^2}\, , \label{5.19}\\
&{}&\cfrac{\partial}{\partial t}\cfrac{1}{|{\bf r}-{\bf r}_p|}=\cfrac{({\bf n}_p{\bf v}_p)}{|{\bf r}-{\bf r}_p|^2}\, ,\label{5.20}
\ea
%%%%%
we can easily seen that the condition \rf{5.17} is satisfied:
%%%%%
\ba{5.19}
&{}&\cfrac{\partial f_0^\beta}{\partial x^\beta}-\cfrac{1}{2}\,\cfrac{\partial h_\mu^\mu}{\partial x^0} =
\cfrac{G_N}{2c^3}\left\{\cfrac{3D-2-\Sigma}{D-2+\Sigma}\sum_pm_p\cfrac{({\bf n}_p{\bf v}_p)}{|{\bf r}-{\bf r}_p|^2}\right.\nn\\
&-&\left.\sum_pm_p\cfrac{({\bf n}_p{\bf v}_p)}{|{\bf r}-{\bf r}_p|^2} -\cfrac{2(D-\Sigma)}{D-2+\Sigma}\sum_pm_p\cfrac{({\bf n}_p{\bf v}_p)}{|{\bf r}-{\bf
r}_p|^2}\right\}=0\, .\nn\\ &{}& \ea
%%%%%

Because the presence of pressure/tension in the extra dimensions requires the uniform smearing of the gravitating masses over the internal space, we provide
the metric coefficients in this case:
%%%%%
\ba{5.20}
g_{00}&\approx& 1+\frac{2\varphi({\bf r})}{c^2}+\frac{2\varphi^2({\bf r})}{c^4}\nn \\
&+&\frac{2G_N^2}{c^4} \sum_p\cfrac{m_p}{|{\bf r}-{\bf r}_p|}\sum_{q\neq p}\frac{m_q}{|{\bf r}_p-{\bf r}_q|} \nn\\
&-&\frac{D-\Sigma}{D-2+\Sigma}\frac{G_N}{c^4}\sum_p\frac{m_pv_p^2}{|{\bf r}-{\bf r}_p|}\, , \ea
%%%%%%
%%%%%%
\ba{5.21} g_{0\alpha}&\approx&\frac{3D-2-\Sigma}{D-2+\Sigma}\frac{G_N}{2c^3}\sum_p\frac{m_p}{|{\bf r}-{\bf r}_p|}\,v^{\alpha}_p \nn\\
&+&\frac{G_N}{2c^3}\sum_p\frac{m_p}{|{\bf r}-{\bf r}_p|}\, n_p^{\alpha} ({\bf n}_p{\bf v}_p)\, , \ea
%%%%%%%
%%%%%%%
\be{5.22}
g_{\alpha\beta}\approx\left(-1+\cfrac{1-\Sigma}{D-2+\Sigma}\cfrac{2\varphi({\bf r})}{c^2}\right) \delta_{\alpha\beta}\, ,
\ee
%%%%%%
%%%%%%
\be{5.23}
g_{\bar\alpha\bar\beta}\approx\left(-1+\cfrac{\omega_{(\bar\alpha-3)}(D-1)+1-
\Sigma}{D-2+\Sigma}\, \cfrac{2\varphi({\bf r})}{c^2}\right)\delta_{\bar\alpha\bar\beta}\, ,
\ee
%%%%%%
where the potential $\varphi({\bf r})$ is given by \rf{5.10}.

Therefore, in this section we have shown that, to be compatible with the gravitational tests, the gravitating masses should be uniformly smeared over the
internal space. This conclusion has the following important effect. Suppose that we have solved for the considered particle the multidimensional quantum
Schr$\mathrm{\ddot{o}}$dinger equation and found its wave function $\Psi (\bf x)$. In general, this function depends on all spatial coordinates ${\bf x}=({\bf
r},{\bf y})$, where ${\bf y}$ are the coordinates in the internal space,  and we can expand it in appropriate eigenfunctions of the compact internal space,
i.e. in the Kaluza-Klein modes. The ground state corresponds to the absence of these particles. In this state the wave function may depend only on the
coordinates ${\bf r}$ of the external space. The classical rest-mass density is proportional to the probability density $|\Psi|^2$. Therefore, the demand that
the rest-mass density depends only on the coordinates of the external space means that the particle can be only in the ground quantum state, and KK excitations
are absent. This looks very unnatural from the point of quantum and statistical physics, because the nonzero temperature must result in excitations.

%%%%%%%%%%%%%%%%%%%%%%%%%%%%%%%%%%%%%%%%%%%%%%%%%%%%%%%%%%%%%%%%%%%%%%%%%%%%%%%%%%%
\section{\label{sec:6}Lagrange function for a many-body system}

Let us construct now the Lagrange function of the many-body system described above. To perform it, we will follow the procedure described in \cite{Landau} (see
\S 106). The Lagrange function of a particle $p$ with the mass $m_p$ in the gravitational field created by the other bodies is given by the expression
%%%%%
\ba{6.1} &{}&L_p=-m_pc\cfrac{ds_p}{dt} \nn\\
&=&-m_p c^2\left(g_{00}+2\sum_\mu g_{0\mu}\cfrac{v^\mu_p}{c}+\sum_{\mu\nu}g_{\mu\nu}\cfrac{v^\mu_pv^\nu_p}{c^2}\right)^{1/2}\, , \ea
%%%%%%
where the metric coefficients are taken at ${\bf r}={\bf r}_p$. We should keep in mind that in the case of the smeared (over the extra dimensions) gravitating
masses, the components of the velocity in the extra dimensions are equal to zero: $v^{\mu}_p = (v_p^{\alpha},v_p^{\bar\alpha}) = (v_p^{\alpha},0)$. It is
convenient to rewrite the metric coefficients \rf{5.20}-\rf{5.22} in the following form:
%%%%%%%
\ba{6} &{}&g_{00}\approx 1+\cfrac{1}{c^2}\,\gamma^{(1)}_{00}+\cfrac{1}{c^4}\,\gamma^{(2)}_{00},\quad g_{0\alpha}\approx\cfrac{1}{c^3}\,\gamma_{0\alpha},\nn\\
&{}&g_{\alpha\beta}\approx\left(-1+\cfrac{1}{c^2}\,\gamma_{(\alpha)}\right)\delta_{\alpha\beta}\, , \nn \ea
%%%%%%
where the meaning of the functions $\gamma$ is evident. Then, we get
%%%%%
\ba{6.2} \cfrac{ds_p}{dt}&\approx& c\left\{1+\cfrac{1}{2c^2}\left[\gamma^{(1)}_{00}-v^2_p\right] \right.\nn\\
&+&\cfrac{1}{2c^4}\left[\gamma^{(2)}_{00}+2\sum_\alpha \gamma_{0\alpha}v^\alpha_p
+\sum_{\alpha\beta}\gamma_{(\alpha)}\delta_{\alpha\beta}v^\alpha_pv^\beta_p \right]\nn \\
&-&\left.\cfrac{1}{8c^4}\left[\gamma^{(1)}_{00}-v^2_p\right]^2\right\}\, .
\ea
%%%%%%
Substituting the explicit form of the metric coefficients \rf{5.20}-\rf{5.22}, we obtain
%%%%%
\ba{6.3}
&{}&L_p = -m_pc^2 +\cfrac{m_pv_p^2}{2}+\cfrac{m_pv_p^4}{8c^2}+G_N\sum_s\cfrac{m_pm_s}{|{\bf r}-{\bf r}_s|}\,\nn \\
&-& \cfrac{1}{2c^2}\,G_N^2\sum_s\sum_q\cfrac{m_pm_sm_q}{|{\bf r}-{\bf r}_s||{\bf r}-{\bf r}_q|}\, \nn\\
&-&\cfrac{1}{c^2}\,G_N^2\sum_s\sum_{q\neq
s}\cfrac{m_pm_sm_q}{|{\bf r}-{\bf r}_s||{\bf r}_s-{\bf r}_q|}\,\nn \\
&+&\cfrac{1}{2c^2}\,G_N\sum_s\cfrac{m_pm_s}{|{\bf r}-{\bf
r}_s|}\left[\mathfrak{a}(D,\Sigma)v_s^2+2\left(\mathfrak{b}(D,\Sigma)+\cfrac{1}{2}\right)v_p^2\right.\nn \\
&-&\left.\mathfrak{c}(D,\Sigma)({\bf v}_s{\bf v}_p)-({\bf n}_s{\bf v}_s)({\bf
n}_s{\bf v}_p)\right]\, .
\ea
%%%%%
Here, we use the following abbreviations:
%%%%%%
\ba{6.4}
\mathfrak{a}(D, \Sigma)&\equiv&\cfrac{D-\Sigma}{D-2+\Sigma}\, ,\quad \mathfrak{b}(D, \Sigma)\equiv\cfrac{1-\Sigma}{D-2+\Sigma}\, ,\nn\\
\mathfrak{c}(D, \Sigma)&\equiv&\cfrac{3D-2-\Sigma}{D-2+\Sigma}\, .
\ea
%%%%%%
We remind that in the expression \rf{6.3} ${\bf r}={\bf r}_p$ and all infinite terms should be cast out. For our purposes, it is sufficient to consider the
case of two particles. Then, for the particle "1", we have the following expression:
%%%%%%
\ba{6.5} L_1&=&\mathrm{f}({\bf v}_1^2)+G_N\cfrac{m_1m_2}{|{\bf r}-{\bf r}_2|}-\cfrac{1}{2c^2}\,G_N^2\cfrac{m_1m_2^2}{|{\bf r}-{\bf r}_2|^2}\nn\\
&-&\frac{1}{c^2}G_N^2
\cfrac{m_1^2m_2}{|{\bf r}-{\bf r}_2||{\bf r}_1-{\bf r}_2|}\nn\\
&+&\cfrac{1}{2c^2}\cfrac{G_Nm_1m_2}{|{\bf r}-{\bf
r}_2|}\left[\mathfrak{a}(D,\Sigma)v_2^2+2\left(\mathfrak{b}(D,\Sigma)+\cfrac{1}{2}\right)v_1^2\right.\nn\\
&-&\left.\mathfrak{c}(D,\Sigma)({\bf v}_1{\bf v}_2)-({\bf n}_2{\bf v}_2)({\bf
n}_2{\bf v}_1)\right]\, ,
\ea
%%%%%%
where $\mathrm{f}({\bf v}_1^2)=m_1v_1^2/2+m_1v_1^4/(8c^2)$ and we drop the term $-m_1 c^2$.

The total Lagrange function of the two-body system should be constructed so that it leads to the correct values of the forces $\left.\partial L_p/\partial {\bf
r}\right|_{{\bf r}={\bf r}_p}$ acting on each of the bodies for given motion of the others \cite{Landau}. To achieve it, we, first, will differentiate $L_1$
with respect to ${\bf r}$, setting ${\bf r}={\bf r}_1$ after that. Then, we should integrate this expression with respect to ${\bf r}_1$. Following this
prescription and taking into account a useful auxiliary relation
%%%%%%
\ba{6.6} &{}&\left(-\cfrac{1}{2c^2}\,G_N^2m_1m_2^2\cfrac{\partial}{\partial{\bf r}}\cfrac{1}{|{\bf r}-{\bf
r}_2|^2}\right.\nn\\
&-&\left.\left.\cfrac{G_N^2}{c^2}\cfrac{m_1^2m_2}{|{\bf r}_1-{\bf r}_2|}
\cfrac{\partial}{\partial{\bf r}}\cfrac{1}{|{\bf r}-{\bf r}_2|}\right)\,\right|_{{\bf r}={\bf r}_1}\nn \\
&=&-\cfrac{1}{2c^2}\,G_N^2m_1m_2(m_1+m_2)\cfrac{\partial}{\partial{\bf r}_1}\cfrac{1}{|{\bf r}_1-{\bf r}_2|^2}\, ,
\ea
%%%%%
we obtain from \rf{6.5} the two-body Lagrange function
%%%%%
\ba{6.7}
L^{(2)}_{1}&=&\mathrm{\tilde f}({\bf v}_1^2, {\bf v}_2^2)+\cfrac{G_Nm_1m_2}{r_{12}}-\cfrac{G_N^2m_1m_2(m_1+m_2)}{2c^2r_{12}^2}\nn\\
&+&\cfrac{G_Nm_1m_2}{2c^2r_{12}}\left[\mathfrak{a}(D,\Sigma)v_2^2+\left(2\mathfrak{b}(D,\Sigma)+1\right)v_1^2\right.\nn\\
&-&\left.\mathfrak{c}(D,\Sigma)({\bf v}_1{\bf v}_2)-({\bf n}_{12}{\bf
v}_1)({\bf n}_{12}{\bf v}_2)\right]\, ,
\ea
%%%%%
where $\mathrm{\tilde f}({\bf v}_1^2, {\bf v}_2^2)=\sum_{a=1}^2m_av_a^2/2+\sum_{a=1}^2m_av_a^4/(8c^2)$ and $r_{12}\equiv |{\bf r}_1-{\bf r}_2|$. It can be
easily seen that $\left.\partial L_1/\partial {\bf r}\right|_{{\bf r}={\bf r}_1} = \partial L^{(2)}_1/\partial {\bf r}_1$. By the same way we can construct the
two-body Lagrange function $L^{(2)}_2$ from the Lagrange function $L_2$ for the particle "2":
%%%%%
\ba{6.8}
L^{(2)}_{2}&=&\mathrm{\tilde f}({\bf v}_1^2, {\bf
v}_2^2)+\cfrac{G_Nm_1m_2}{r_{12}}-\cfrac{G_N^2m_1m_2(m_1+m_2)}{2c^2r_{12}^2}\nn\\
&+&\cfrac{G_Nm_1m_2}{2c^2r_{12}}\left[\mathfrak{a}(D,\Sigma)v_1^2+\left(2\mathfrak{b}(D,\Sigma)+1\right)v_2^2\right.\nn\\
&-&\left.\mathfrak{c}(D,\Sigma)({\bf v}_1{\bf v}_2)-({\bf n}_{12}{\bf
v}_1)({\bf n}_{12}{\bf v}_2)\right]\, .
\ea
%%%%%
It is worth noting that both $L^{(2)}_{1}$ and $L^{(2)}_{2}$ are reduced to the Lagrange function of the two-body system in \cite{Landau} if we assume that $D=3$,
$\Sigma=0$.

Obviously, the Lagrange functions $L^{(2)}_{1}$ and $L^{(2)}_{2}$ should be symmetric with respect to permutations of particles 1 and 2 and should coincide
with each other. This requires the following condition:
%%%%%%
\be{6.11}
\mathfrak{a}(D,\Sigma)=2\mathfrak{b}(D,\Sigma)+1\, ,
\ee
%%%%%%
which is satisfied identically for any value of $\Sigma$. Therefore, we construct the two-body Lagrange function for any value of the parameters of the
equation of state in the extra dimensions.

%%%%%%%%%%%%%%%%%%%%%%%%%%%%%%%%%%%%%%%%%%%%%%%%%%%%%%%%%%%%%%%%%%%%%%%%%%%%%%%%%%%%%%%%%%%%%%%%%%%%%%%%%%%%%%%%%%%
%%%%%%%%%%%%%%%%%%%%%%%%%%%%%%%%%%%%%%%%%%%%%%%%%%%%%%%%%%%%%%%%%%%%%%%%%%%%%%%%%%%%%%%%%%%%%%%%%%%%%%%%%%%%%%%%%%%

\vspace{0.3cm}

\section{Gravitational tests}

%%%%%%
It can be easily seen that the components of the metrics coefficients in the external/our space \rf{5.20}-\rf{5.22} as well as the two-body Lagrange functions
\rf{6.7} and \rf{6.8} exactly coincide with the corresponding expressions in General Relativity for the value $\Sigma =
\sum_{\bar\alpha}\omega_{(\bar\alpha-3)} = - (D-3)/2$, i.e. the latent soliton case \cite{EZ5}. Black strings and black branes are particular cases of it.
Therefore, the known gravitational tests in this case give the same results as for General Relativity. In other words, we get a good agreement with
observations. It is of interest to obtain an experimental restriction on deviation from this value. For this purpose, we write $\Sigma$ in the following form:
%%%%%%
\be{26}
\Sigma = -\frac{D-3}{2}+\varepsilon
\ee
%%%%%%
and find the experimental limitations on $\varepsilon$.

\

{\it PPN parameters}

\vspace{0.2cm}

To get the parameterized post-Newtonian parameters (PPN) $\beta$ and $\gamma$, we consider the case of one particle at rest. Then, we can easily obtain from
Eqs. \rf{5.20} and \rf{5.22} that
%%%%%
\be{27}
\beta =1\, ,\quad \gamma = \frac{1-\Sigma}{D-2+\Sigma}\, ,
\ee
%%%%%
i.e. PPN parameter $\beta$ exactly coincides with the value in the General Relativity.
There are strong experimental restrictions on the value of $\gamma$. The tightest constraint on $\gamma$ comes from  the Shapiro time-delay experiment using
the Cassini spacecraft, namely: $\gamma-1 =(2.1\pm 2.3)\times 10^{-5}$ \cite{Will2,JKh,Bertotti}. In our case
%%%%%
\be{28}
\gamma - 1 \approx -\frac{4\varepsilon}{D-1}\, .
\ee
%%%%%
Therefore, the Shapiro time-delay experiment results in the following limitation:
%%%%%
\be{29}
|\varepsilon| \lesssim \frac{D-1}{2}\times 10^{-5}\, .
\ee
%%%%%
%%%%%%%%%%%%%%%%%%%%%%%%%%%%%%%%%%%%%%%%%%%%%%%%%%%%%%%%%%%%%%%%%%%%%%%%%%%%%%%%%%%%%%%%%%%%%%%%%%%%%

\

{\it Perihelion shift of the Mercury}

\vspace{0.2cm}

For a test body orbiting around the gravitating mass $m$, the perihelion shift for one period is given by the formula \cite{Will2,Will}
%%%%%
\be{30}
\delta \psi = \frac13 \left(2+2\gamma -\beta\right)\, \frac{6\pi G_N m}{c^2a(1-e^2)}\equiv
\frac13 \left(2+2\gamma -\beta\right)\, \delta \psi_{GR}\, ,
\ee
%%%%%%%
with $a$ and $e$ being the semi-major axis and the eccentricity of the ellipse, respectively. $\delta \psi_{GR}$ is the value for General Relativity. In the
case of Mercury this calculated value is equal to 42.98 arcsec per century \cite{Will2,Will3}. This predicted relativistic advance agrees with the observations
to about 0.1\% \cite{Will2}. Substituting the PPN parameters \rf{27} in this formula, we obtain the advance in our case:
%%%%
\be{31}
\delta \psi = \frac13 \frac{D-\Sigma}{D-2+\Sigma}\, \delta \psi_{GR} \approx \left(1-\frac{8}{3(D-1)}\varepsilon\right)\delta \psi_{GR}\, .
\ee
%%%%
Obviously, to be in agreement with the observation no worse than General Relativity, the parameter $\varepsilon$ should satisfy
the condition
%%%%%%%
\be{32}
|\varepsilon| \lesssim \frac{3(D-1)}{8}\times 10^{-3}\, .
\ee
%%%%%%%
Therefore, this limitation is less strong than \rf{29}.

%%%%%%%%%%%%%%%%%%%%%%%%%%%%%%%%%%%%%%%%%%%%%%%%%%%%%%%%%%%%%%%%%%%%%%%%%%%%%%%%%%%%%%%%%%%%%%%

\

{\it Periastron shift of the relativistic binary pulsar
PSR B1913+16}

\vspace{0.2cm}

Much more strong limitation can be found from the measurement of the periastron shift of the relativistic binary pulsar. First, the advance of periastron in
these systems in many orders of magnitude bigger than for the Mercury. Second, the measurements are extremely accurate. For example, for the pulsar PSR
B1913+16 the shift is $4.226598\pm 0.000005$ degree per year \cite{pulsar}. For such system both the pulsar and companion have comparable masses. In the case
of General Relativity, a solution for orbital parameters yields mass estimates for the pulsar and its companion, $m_1 = 1.4398\pm 0.0002 M_{\odot}$ and $m_2 =
1.3886 \pm 0.0002 M_{\odot}$, respectively. It is worth noting that these are calculated values (not observable!) which are valid for General Relativity.
Because two bodies have comparable masses (and one of them cannot be considered as a test body), to get a formula for the advance we need a two-body
Lagrangian. Then, following the problem 3 in \S 106 \cite{Landau} we get for our two-body Lagrangians \rf{6.7} and \rf{6.8} the desired formula in the form of
\rf{31} with the well known General Relativity expression
%%%%%%
\be{33}
\delta \psi_{GR} = \frac{6\pi G_N (m_1+m_2)}{c^2a(1-e^2)}\, .
\ee
%%%%%%
In future, independent measurements of masses $m_1$ and $m_2$ will allow us to obtain a high accuracy restriction on parameter $\varepsilon$.

%%%%%%%%%%%%%%%%%%%%%%%%%%%%%%%%%%%%%%%%%%%%%%%%%%%%%%%%%%%%%%%%%%%%%%%%%%%%%%%%%%%%%%%%%%%%%%%%%%%%%%%%%%%%%%%%%
%%%%%%%%%%%%%%%%%%%%%%%%%%%%%%%%%%%%%%%%%%%%%%%%%%%%%%%%%%%%%%%%%%%%%%%%%%%%%%%%%%%%%%%%%%%%%%%%%%%%%%%%%%%%%%%%%%

\vspace{0.3cm}

\section{Summary}

In this paper, we have constructed the Lagrange function for a two-body system in the case of Kaluza-Klein models with toroidal compactification of the extra
dimensions. The case of more than two bodies is straightforward. We supposed that gravitating bodies are pressureless in the external/our space. This is a
natural approximation for ordinary astrophysical objects such as our Sun. For example, this approach works well for calculating the gravitational experiments
in the Solar system \cite{Landau}. In the case of pulsars, pressure is not small but still much less than the energy density. Hence, the pressureless approach
is used in General Relativity to get the formula \rf{33} which is in very good agreement with the observations of advance of periastron of the pulsar PSR
B1913+16.

With respect to the internal space, we supposed that  gravitating masses may have nonzero parameters $\omega_{(\bar\alpha -3)} \, (\bar\alpha =4,\ldots ,D) $
of the equations of state  in the extra dimensions. We have shown that the Lagrange function of this many-body system can be constructed for any value of the
parameter $\Sigma =\sum_{\bar\alpha} \omega_{(\bar\alpha -3)}$.

To construct the many-body Lagrangian, as well as to get the formulas for the gravitational tests, we obtained the  metrics components $g_{00}$ up to
$O(1/c^4)$, $g_{0\alpha}$ up to $O(1/c^3)$ and $g_{\alpha\beta}$ up to $O(1/c^2)$. These expressions exactly coincide with the corresponding formulas in
General Relativity for the value $\Sigma = \sum_{\bar\alpha}\omega_{(\bar\alpha-3)} = - (D-3)/2$. This is the latent soliton case \cite{EZ5}. Black
strings/branes are particular cases of it with all $\omega_{(\bar\alpha -3)}=-1/2\, \forall \bar\alpha$. Obviously, the known gravitational tests (PPN
parameters, perihelion/periastron shift) in this case give the same results as for General Relativity. On the other hand, we used these tests to get the
restrictions on the deviation from the latent soliton value. At the present, the most strong restriction follows from the time delay of radar echoes (the
Cassini spacecraft mission). The two-body Lagrange function allowed us to get the formula for the advance of the periastron. In future, when the masses of the
binary pulsar system PSR B1913+16 will be measured (rather than calculated using the formula of General Relativity), the advance of this periastron can be used
to get the restriction with very high accuracy. All obtained limitations indicate very small deviation from the latent soliton value. Therefore, the
pressureless case $\Sigma =0$ in the internal space is forbidden, in full agreement with the results of the paper \cite{EZ3}.  This conclusion does not depend
on the size of extra dimensions. The physical reason of it is that in the case of toroidal compactification, only in the case of latent solitons the variations
of the total volume of the internal space are absent \cite{EZ6}.

One more important result obtained in this paper is worth noting. As we have shown above (see also \cite{EZ4,EZ5,EZ6}), tension in the internal spaces is the
necessary condition to satisfy the gravitational experiments in KK models with toroidal compactification. In our paper, we have proven that the presence of
pressure/tension in the internal space leads necessarily to the uniform smearing of the gravitating masses over the internal space. For example, black
strings/branes have tension in the internal space (see, e.g., \cite{Traschen}). Therefore, they should be smeared. However, uniformly smeared gravitating
bodies cannot have  excited KK states (KK particles), which looks unnatural from the point of quantum mechanics and statistical physics. In our opinion, this
is a big disadvantage of the Kaluza-Klein models with the toroidal compactification. It is of interest to check this property for models with other types of
compactification (e.g. Ricci-flat, spherical). This is the subject of our subsequent study.
%%%%%%%%%%%%%%%%%%%%%%%%%%%%%%%%%%%%%%%%%%%%%%%%%%%%%%%%%%%%%%%%%%%%%%%%%%%

%\vspace{0.5cm}

\section*{Acknowledgments}

\ \ \ \ \ This work was supported in part by the "Cosmomicro-physics-2" programme of the Physics and Astronomy Division of the National Academy of Sciences of
Ukraine. The work of M. Eingorn was supported by NSF CREST award HRD-0833184 and NASA grant NNX09AV07A.

%%%%%%%%%%%%%%%%%%%%%%%%%%%%%%%%%%%%%%%%%%%%%%%%%%%%%%%%%%%%%%%%%%%%%%%%%%%%%%%%%%%%%%%%%%%%%%%%%%%%%%%%%%%%%%%%%
%%%%%%%%%%%%%%%%%%%%%%%%%%%%%%%%%%%%%%%%%%%%%%%%%%%%%%%%%%%%%%%%%%%%%%%%%%%%%%%%

\end{document}